# Towards Interoperability Testing of Smart Energy Systems – An Overview and Discussion of Possibilities

*Thomas I. Strasser[1*], Edmund Widl[1], René A. Kuchenbuch[2], Laura Lázaro-Elorriaga[3], Borja Tellado Laraudogoitia[3], Mirko Ginocchi[4], Thanakorn Penthong[4], Ferdinanda Ponci[4], Amelie Gyrard[5], Antonio Kung[5], Carlos A. Mac Gregor[6], Carmen Garcia Montero[7], Eduardo Relano Algaba[7]*

[1]*Center for Energy, AIT Austrian Institute of Technology, Vienna, Austria*
[2]*Division Energy, OFFIS e.V., Oldenburg, Germany*
[3]*Energy, Climate and Urban Transition, Fundacion Tecnalia Research & Innovation, Derio, Spain*
[4]*Institute for Automation of Complex Power Systems, RWTH Aachen University, Aachen, Germany*
[5]*Trialog, Paris, France*
[6]*B.A.U.M. Consult, Munich, Germany*
[7]*European Network of Transmission System Operators for Electricity (ENTSO-E), Brussels, Belgium*
*\*thomas.strasser@ait.ac.at*

## Abstract

Interoperability is the key to implementing a wide range of energy systems applications. It involves the seamless cooperation of different methods and components. With smart energy systems, interoperability faces challenges due to integrating different approaches and technologies. This includes dealing with heterogeneous approaches with various communication protocols and data formats. However, it is essential for smart energy systems to carry out thorough interoperability tests. They are usually diverse, and challenging, thus requiring careful consideration of compatibility issues and complex integration scenarios. Overcoming these challenges requires a systematic approach that includes thorough test planning, rigorous testing, and continuous test monitoring. Although numerous testing approaches exist, most are more developed at the component/device level than at the system level. Consequently, there are few approaches and related facilities to test the interoperability of smart energy approaches and solutions at the system level. This work analyses existing interoperability test concepts, identifies enablers and the potential for harmonisation of procedures, and proposes further developments of these approaches.

## 1 Introduction

Interoperability is a crucial component for the successful integration of diverse applications and services within power and energy systems [1], [2]. This becomes particularly complex in smart energy applications where multiple domains, such as electric power, control systems, and Information and Communication Technology (ICT), need to seamlessly work together. This necessitates the smooth collaboration of various methodologies, solutions, components, and devices. Several strategies can be employed to achieve interoperability among these interconnected systems, including [3], [4]:

- *Requirements engineering:* Adequate requirements engineering helps to understand stakeholder needs and supports therefore the identification of interoperability needs.
- *Interoperability-by-design:* Incorporating interoperability considerations from the design phase.
- *Application of reference architectures:* Offering blueprints that define the structure and interaction of systems and components.
- *Compliance with standards:* Following established rules, guidelines, and standards.

However, following these principles alone does not guarantee interoperability in smart energy systems. Therefore, it is crucial to perform comprehensive interoperability tests, issue test reports, and provide certificates (or labels) that confirm the successful completion of these tests.

The field of interoperability testing is marked by a variety of procedures, categorisations, evaluations, and assessment criteria, mirroring the diverse interests across different initiatives. Currently, there is a gap in the comprehensive synthesis of interoperability testing methodologies and approaches applied to power and energy systems by related testing facilities and laboratories across Europe.

While there are numerous testing approaches and methods, they are often more detailed at the component/device level than at the system level, which is still largely in the research and development stage. Moreover, the existing concepts, related procedures, and assessment criteria are diverse and cater to a wide range of interests. Therefore, harmonisation of these approaches is necessary. Such an effort is currently underway in the European int:net project [4].

This work is a result of the aforementioned project, and it analyses existing test concepts and procedures, identifies opportunities for harmonisation, and suggests further advancements in these approaches and methods.

The rest of this paper is structured as follows: Section 2 presents the methodology as well as selected applications and interoperability needs used for this work. Section 3 identifies challenges and obstacles for interoperability testing, while Section 4 derives potential enablers. Section 5 discusses best practices and recommendations for harmonisation. Finally, Section 6 concludes this work.

## 2 Applied Methodology

### 2.1 Key Aspects

To identify challenges related to interoperability testing of smart energy systems and potential areas for harmonisation, a comprehensive four-step approach is employed. In the first step, a variety of testing initiatives, approaches, applications, and use cases from the targeted domain (i.e., power and energy systems) is gathered. Mainly project-related resources are being used for this step [6]-[8]. In the second step (cf. Section 3), a thorough analysis of these initiatives and approaches is conducted. The challenges are then discussed under three key areas of interest for interoperability testing:

- *Applications:* This pertains to the question, "What needs to be covered by interoperability testing?"
- *Infrastructures:* This involves determining the virtual and/or physical infrastructures and facilities required for interoperability testing, with a particular focus on the implementation of interoperability testing setups.
- *Procedures:* This concerns the question, "How should interoperability testing be carried out?" with a focus on both the specification and execution of interoperability tests.

In the third step (cf. Section 4), a set of enablers is derived to address these challenges. Also, recommendations for the harmonisation and further development of these approaches are being made. Finally, in the fourth step, the outcome of the analysis is documented. This step not only provides a comprehensive record of the main findings but also serves as a valuable resource for future research and development activities in the field of interoperability testing.

### 2.2 Selected Applications and Interoperability Needs

In this work, the following six different applications in connection with intelligent energy systems are analysed:

1) *Network of Electric Vehicle (EV) Charging Stations:* Interoperability enables seamless interaction among diverse EV charging stations. However, the lack of standardised protocols leads to compatibility issues, hindering a user-friendly EV charging network. Interoperability, through standardised protocols, could enhance the EV charging experience and promote wider EV adoption.
2) *Supervisory Control and Data Acquisition (SCADA) Systems,* used in industrial control, face challenges including cyber-security, compatibility, scalability, reliability, maintenance, and interoperability. Diverse technologies and protocols can hinder effective communication, making these systems often inflexible and centralised.
3) *Digital Substations,* designed based on the IEC 61850 standards family [9], promote interoperability among devices from various manufacturers. The use of a common data model and description language (i.e., SCL), as well as protocols (i.e., GOOSE, MMS, SV, etc.), ensures standardised information representation, enabling seamless information exchange.
4) *Energy Smart Appliances* are devices that can control their energy consumption and respond to external stimuli, contributing significantly to household demand-side flexibility. However, ensuring interoperability among multi-vendor solutions is challenging due to the need for compliance with rapidly evolving standards [10].
5) *Energy Data Spaces:* Applications complying with initiatives like the International Data Spaces Association (IDSA) and Gaia-X prioritize data sovereignty and robust security measures to ensure flexibility and interoperability without compromising data security. Adherence to common data formats, protocols, architectures and Application Programming Interfaces (APIs), along with transparent and accountable data handling practices and clear data governance policies, builds trust with stakeholders and promotes the growth of the data space ecosystem [8].
6) *Common Grid Model Data Exchange and Services:* Transmission System Operators (TSOs) are required to ensure the security and availability of the electrical power grid. The Common Grid Model Exchange Specification (CGMES), based on the IEC 61970 standards family (i.e., CIM) [11], an enterprise standard used for network data exchange, aids in merging individual TSO network models into a pan-European one, enabling various Regional Coordination Centre (RCC) services. The successful implementation of this standardised data exchange involves the main parties in the adoption process, including ENTSO-E member TSOs, ENTSO-E as a centralised standardisation organisation, and software implementers.

The selection of the above applications is based on previous int.net results [7]. Each of them presents unique challenges and needs for interoperability testing, which are discussed below.

## 3 Challenges for Interoperability Testing

Hereafter, an overview of interoperability testing challenges related to the three key areas of interest (cf. Section 2.1) is provided.

### 3.1 Application-related Challenges

Table 1 lists the identified main challenges and requirements for interoperability testing from an application point of view.

Table 1 Application-related interoperability testing challenges and requirements (adopted from [12])

| Challenges | Requirements |
|---|---|
| *Network of EV Charging Stations* | |
| • Fragmentation of networks<br>• Complex user experience<br>• Roaming problems | • Standardised protocols, data models, and interface specifications<br>• Common principles for data sharing<br>• Compatibility<br>• Seamless access |

| Challenges | Requirements |
|---|---|
| *SCADA Systems* | |
| • Interop. among heterogenous SCADA syst. & emerging techn.<br>• Integration with legacy systems | • Standardised protocols, data models, and interfaces with mandatory specifications |
| *Digital Substations* | |
| • Interoperability among different vendor solutions | • Standardised protocols, data models, and interface specifications |
| *Energy Smart Appliances* | |
| • Developing regulatory rules<br>• Rapid technology developments | • Common principles for data sharing<br>• Agreed stakeholder commitments |
| *Energy Data Spaces* | |
| • Common data ecosystem (specifications, rules, etc.) | • Privacy, trust, and data protection<br>• Lawfulness, fairness, and transparency<br>• Purpose limitation and data minimisation<br>• Accuracy and integrity |
| *Common Grid Model Data Exchange and Services* | |
| • Culture change adoption<br>• Differentiation of interoperability<br>• Documentation and communication of use cases | • CGMES developing and testing<br>• Version control and release management<br>• Validation and acceptance testing |

*3.2 Infrastructure-related Challenges*

The discovered main challenges and requirements from an interoperability testing infrastructure point of view are described in Table 2.

Table 2 Infrastructure-related interoperability testing challenges and requirements (adopted from [12])

| Challenges | Requirements |
|---|---|
| *Network of EV Charging Stations* | |
| • Ultra-fast charging<br>• Wireless charging<br>• Vehicle-to-grid paradigm | • Standardised charging ports and connectors<br>• Common control and communication technology for EV charging<br>• Integr. with power grids/networks |
| *SCADA Systems* | |
| • Harmon. comm. approaches<br>• Legacy integration<br>• Cloud-edge approaches integration | • Standardized information models<br>• Flexible communication architecture<br>• Safety and integrity |
| *Digital Substations* | |
| • SCL interoperability<br>• Real-time platform interoperability for SV and GOOSE handling<br>• Device virtualisation | • Standardized SCL interpretation<br>• Integr. strategies for legacy systems<br>• Enhanced security measures<br>• Time synchronisation |
| *Energy Smart Appliances* | |
| • Harmon. of proprietary solutions<br>• Harmonized/standardised use cases and information exchange | • Syntactic & semantic interop.<br>• Robust security mechanisms<br>• Support for communication standards and interfaces |
| *Energy Data Spaces* | |
| • Trust sovereign identities<br>• Different concepts of the world | • Data sovereignty and decentralisation<br>• Data modelling and interfacing |
| *Common Grid Model Data Exchange and Services* | |
| • Reference data development and implementation<br>• Test data validation | • Cutting-edge semantic technology support<br>• Collaborative platforms |

*3.3 Procedure-related Challenges*

Besides the above-outlined application and infrastructure-related interoperability testing challenges and requirements, there exists a significant number of different procedures related to four main testing-related phases. They can be summarised as (i) test case identification and definition (incl. profiling), (ii) test planning and test set-up specification, (iii) test execution, and (iv) test reporting. The following approaches have been selected and analysed in detail [12]:

- *Methodologies:* JRC Smart Grid Interoperability Testing Methodology, European Code of Conduct (CoC) for Energy Smart Appliances (incl. related interoperability testing), SMARTGRIDS Austria IES-Process (i.e., adopted interoperability testing approach from healthcare domain), and ERIGrid Holistic Test Description (i.e., system-level test description approach for smart grid applications).

- *Framework and tools:* JRC Smart Grid Design of Interoperability Tests (SG-DoIT, linked with the JRC interoperability testing approach), EU DIGIT Interoperability Test Bed (i.e., generic tool for ICT interoperability testing), IHE Gazelle Open-source Platform for Test Management (i.e., interoperability testing framework and related tool for healthcare applications), ENTSO-E CGMES Conformity Assessment Framework (i.e., testing framework for TSOs based on CIM), (NIST) Framework and Roadmap for Smart Grid Interoperability Standards, int:net EMINENT interoperability maturity model (i.e., framework and related tool for assessing and improving interoperability efforts), and AIT VLab (i.e., interoperability-by-design framework).

- *Standards, guidelines, and policies:* ISO/IEC 25000 standards series (i.e., SQuaRE standards family related to product quality), ISO/IEC 30194 (i.e., related to the Internet of Things (IoT) and digital twins), ISO/IEC 27564 (i.e., related to privacy engineering), and ISO/IEC 2182 (i.e., related to behavioural and policy interoperability).

- *Other approaches from the literature:* Metamodel for IoT Testing, Design of Experiments (DoE) in the Methodology for Interoperability Testing, Semantic Web Best Practices applied to Energy Ontologies, and EEBUS Living Lab Cologne.

Table 3 depicts the main challenges and requirements for interoperability testing procedures.

Table 3 Procedures-related interoperability testing challenges and requirements (adopted from [12])

| Challenges | Requirements |
|---|---|
| *Methodologies* | |
| • Diversity of testing methodologies | • Provision of harmonised testing procedures<br>• Provision of a test facility ontology |
| • Approaches usually developed for a specific purpose/area (transmission/distribution grid, building, utility automaton, ESA, etc.)<br>• Diversity in interoperability profiling and test case specification | • Organisation of periodic interoperability testing events<br>• Exchange and sharing of knowledge/experiences during testing events |
| *Frameworks and Tools* | |
| • Diversity of testing frameworks and tools<br>• Approaches usually developed for a specific purpose/area (transmission/distribution grid, utility automaton, etc.)<br>• Diversity in interoperability profiling and test case specification<br>• Diversity of models to test | • Provision of harmonised technical frameworks<br>• Provision of supporting tools for interoperability testing<br>• Provision of a test facility ontology<br>• Publication of interop. profiles<br>• Publication of test case specif.<br>• Provision of a blueprint for interop.-compliant testing facilities<br>• Provision of a test facility ontology<br>• Sharing of best practices btw. the interoperability testing facilities |
| *Standards, Guidelines, and Policies* | |
| • Ontology diversity (CIM, IEC 61850, SAREF, etc.)<br>• Diversity of semantic representations of models to test<br>• Governance of testing (interaction of heterogeneous supply chains)<br>• Testing compliance with policies<br>• Diversity in interoperability labelling and certification | • Provision of a common energy ontology based on domain approaches (CIM, IEC 61850, SAREF, etc.) or integration approaches for harmonisation<br>• Establishment of a Pan-European community of interoperability test facilities<br>• Establishment of internist. conversations between the interop. test facilities for info exchange and harmonisation of procedures<br>• Provision of an interoperability-compliant/approved label<br>• Creation of interoperability certification centres<br>• Provision of interoperability test facility qualification procedures |
| *Other Approaches from the Literature* | |
| • Diversity in interoperability profiling and test case specification | • Adoption of best practices from other domains<br>• Sharing of lessons learned and experiences with interoperability testing |
| *Communication and Adoption Challenges* | |
| • Assess and communicate CGMES v3.0 benefits<br>• Enhance bidirectional communication among stakeholders<br>• Improve organization and scope of interoperability tests<br>• Plan interoperability tests with sufficient preparation time<br>• Conduct transparent public technical discussions | • Early announcement (close to one year in advance) and organisation of interoperability testing events<br>• Ensure business needs are transmitted in the form of use cases. |
| *Conformity Assessment Challenges* | |
| • Encourage active TSO/ DSO participation<br>• Automate conformity assessment<br>• Optimize resources with technological solutions<br>• Enhance data exchange quality and efficiency<br>• Minimize manual intervention and improve accuracy | • Definition of business quality document not conflicting with standards |

## 4 Enablers for Interoperability Testing

Hereafter, an overview of enablers for interoperability testing related to the three key areas of interest (cf. Section 2.1) is provided.

### 4.1 Enablers for Applications

Interoperability testing in smart energy applications is enabled by several key factors.
EV Charging Stations are being developed with common protocols and data formats for seamless communication between stations and management systems, thanks to organisations like the Open Charge Alliance (OCA) and the OpenADR Alliance. Smart charging infrastructure implementation enables features such as real-time data sharing, dynamic load management, and automated billing. Bodies like the International Electrotechnical Commission (IEC) and OCA steer rigorous testing and certification processes. Roaming platforms serve as intermediaries among charging network operators, facilitating integrated billing, authentication, and roaming services for EV drivers.
Addressing interoperability issues among different SCADA Systems can be achieved using standardised communication protocols and middleware concepts. This enhances and facilitates inter-system communication, with standardisation bodies and organisations ensuring effective operation in complex industrial environments.
Digital Substations rely on a set of IEC 61850 standards to validate device data model compliance, ensuring proper device operation. A test procedure is necessary to confirm the correct operation of all control, monitoring, and protection functions before implementation.
For Energy Smart Appliances, it is essential to prepare a set of basic principles for all stakeholders involved in manufacturing. The CoC for the Interoperability of Energy Smart Appliances could increase the number of interoperable solutions in the European market [16]. Enough CoC signatories would guarantee the convergence of a reference ontology framework and the adoption of open standards.
Energy Data Spaces are guided and framed by initiatives like IDSA, Gaia-X, and Data Spaces Business Alliance (DSBA) for organizations to build compliant and interoperable data ecosystems. These initiatives provide standards, principles, and architectures to facilitate secure, interoperable, and trustworthy data exchange ecosystems. Gaia-X emphasizes federated data infrastructure, promoting data sovereignty and transparency across Europe. DSBA aims to harmonize data governance regulations within the European Union, while IDSA focuses on secure data exchange within international data spaces.

### 4.2 Enablers for Infrastructure

Interoperability testing facilities rely on several key enablers. Standardized protocols and data modelling standards allow seamless communication and integration across different systems and devices. Open standards and protocols, such as IEC 61850, Modbus, and DNP3, foster deployment and communication between diverse elements of the smart grid infrastructure. Interoperability platforms and middleware concepts facil-itate smooth data and service exchange among disparate systems, applications, and devices, enhancing scalability and flexibility. Deploying interoperable connectors and charging systems for EVs ensures a smooth charging experience for EV owners. The development of a reference ontology, such as SAREF and SAREF4ENER, is a key enabler of IoT semantic interoperability. Lastly, the development of a common European data space for testing facilities aids in the discovery of interoperability testing possibilities and capabilities.

*4.3 Enablers for Procedures*

Several enablers for procedures have been identified and are displayed in Table 4. These enablers, which are domain-independent and presented from a general interoperability perspective, also include those specific to smart energy systems.

Table 4 Overview of enablers for procedures (adopted from [12])

| Procedures | Generic | Energy Specific |
|---|---|---|
| *Technical* | • Common test bed architecture | • Common test bed solution |
| *Semantical* | • Procedure to construct testable models based on ontologies and procedure to generate test cases<br>• Procedure to define testable profiles and test cases | • Procedure to test encapsulation (e.g., not using some terms in cross-domain interoperability) |
| *Organisational* | • Domain testing methodology<br>• Domain connectathon<br>• Cross-domain testing method.<br>• Cross-domain connetathon | • Specific domain standardised profiles and models |
| *Policy* | • Dynamic testing of policy<br>• Dynamic verif. of secure logs<br>• Static verification of secure logs | • Domain-specific practices |

*4.4 Summary of Enablers*

Table 5 summarises the above findings.

Table 5 Summary of interoperability testing enablers (adopted from [12])

| Application | Infrastructure | Procedures |
|---|---|---|
| *Technical* | | |
| • Open technology stacks | • Open standards for Hardware-in-the-Loop (HIL), digital twins, and data spaces<br>• Laboratory dataspaces | • Testing frameworks |
| *Semantical* | | |
| • Open interface stand.<br>• Enablers to develop testable models (patterns) | • Open domain ontologies<br>• Testable models (pattern) | • Common IoT profiles<br>• Inventory ontology<br>• Testable models (pattern repository) |
| *Organisational* | | |
| • System-level functional testing<br>• User acceptance<br>• Enablers to develop testable interaction & behaviour models | • Testing community<br>• Cybersecurity and trustworthiness support | • Testing methodology<br>• Organisation of testing events (e.g., connectathons, hackathons, plug-fests) |
| *Policy* | | |
| • Enablers to develop testable application policies | • Testable models policy implementation<br>• Testable models for data usage enforcem. | • Labelling |
| *Legal* | | |
| • Regulatory sandboxes | • Com. IoT profile evaluation methodology<br>• Connectors evaluation methodology<br>• Pattern eval. method. | • Certification |

# 5 Discussion

Based on the main findings from the previous Sections 3 and 4, the following best practices and recommendations are worth mentioning.

*5.1 Best Practices*

ENTSO-E provides a Conformity Assessment Scheme for software tools working with CGMES models in the Common Grid Model (CGM) building process [13]. This framework offers detailed guidelines for conformity and interoperability of CIM-based models, such as use cases, test configurations, and application profiles. Vendors can test their understanding of the specifications or standards the framework is based on in the standard-vetting interoperability test (SV-IOP) test. The outcomes of these events are summarised and published, helping to improve the CGMES standard [14].

The JRC has defined a general interoperability testing methodology for smart grid applications and developed related tools like the SG-DoIT [15]. Additionally, a specific testing facility, the SGILab, has been realized for interoperability testing. Furthermore, JRC has further developed this interoperability testing methodology for usage in energy smart appliances together with the provision of a CoC [16]. Wide adoption and use of this approach in the power and energy systems domain would be highly beneficial, especially for IoT-based solutions.

EU DIGIT provides an interoperability test bed for ICT systems that can also be used for smart energy systems applications to test their conformity to particular specifications. [17].

The SMARTGRIDS Austria technology platform adopted the IHE interoperability testing framework from the healthcare sector for smart grid applications [18]. This, along with the usage of the test management tool Gazelle and the organisation of regular connectathon events, forms a powerful interoperability testing approach for organising and executing peer-to-peer interoperability tests.

EEBUS provides a living laboratory for the interoperability assessment of their protocol suite for IoT devices [19]. This laboratory is used by the members of the association for common interoperability tests. Also, regular test events are being planned to improve the interoperability of EEBUS-compliant solutions as well as the knowledge exchange between the EEBUS members.

The int:net project provides an approach called EMINENT and the related tool for accessing and measuring interoperability maturity. To achieve certain maturity levels, the maturity model envisions the establishment of processes for interoperability testing, whereby the maturity model can also provide guidelines for wide-ranging topics (e.g., governance) and therefore its adoption for interoperability testing approaches and facilities seems promising [4].

ERIGrid developed a system-level testing approach and corresponding tools/templates for test cases, test specifications, and experiment specification descriptions [20]. Also, a public repository for the publication and sharing of such descriptions is provided. Moreover, in the successor project, ERIGrid 2.0, a profiling approach for test cases and their categorisation is introduced [21]. Similar to the JRC interoperability testing approach, statistical DoE-based planning of interoperability experiments is suggested.

*5.2 Recommendations for Harmonisation*

In terms of application, the focus is on utilising open technology stacks, open interface standards, and testable models. Emphasis is placed on system-level functional testing, user acceptance, and conformance testing. Additionally, the provision of testable application policies and regulatory sandboxes is integral to the process.

In terms of infrastructure, the use of testing facility/laboratory data spaces and open standards for HIL, digital twins, and data spaces is suggested. The use of domain ontologies and the formation of testing communities for knowledge exchange are also important. Additionally, the provision of cybersecurity and trustworthiness support, certification/labelling of testing infrastructures, and the realisation of test facilities and living labs, such as JRC SGILab and EEBUS Living Lab Cologne, are key components of this approach.

In terms of procedures, the approach involves the use of domain standards like IEC 62559 for use case definition linked with the Smart Grid Architecture Model (SGAM). It includes the definition of interoperability profiles according to the JRC Interoperability Testing Method and IES-Process, and the specification and provision of test cases, test specifications, and experiment specifications in line with the ERIGrid Holistic Test Description approach. Additionally, testing methodologies and frameworks are provided according to ENTSO-E CGMES Conformity Assessment, JRC Energy Smart Appliances CoC, or SMARTGRIDS Austria IES-Process.

# 6 Conclusions

To summarise, interoperability testing is essential for interoperable solutions in the power and energy domain, especially for smart energy systems solutions. However, the landscape of existing testing approaches, concepts, and procedures is diverse. Therefore, a harmonisation of them makes sense. This work identifies the main challenges and enablers to overcome the existing obstacles. Therefore, a comprehensive analysis has been conducted and summarized in this work considering the application, infrastructure, and procedures. Based on it, several recommendations have been made for further development and harmonisation.

Finally, more details about the main findings of this work are provided in the related int:net report [12].

# 7 Acknowledgements

This work has received funding from the European Union's Horizon Europe research and innovation programme under grant agreement N°101070086 (int:net).

# 8 References

[1] Ayadi, F., Colak, I., and Bayindir, R.: 'Interoperability in Smart Grid', 7th Int. Conference on Smart Grid (icSmartGrid), Newcastle, NSW, Australia, 2019, pp. 165–169

[2] Rana, B, Singh, Y., and Singh, P.K.: 'A systematic survey on internet of things: Energy efficiency and interoperability perspective', Transactions on Emerging Telecommunications Technologies, 2021, 32(8), e4166

[3] Reif, V., Strasser, T.I., Jimeno, J., et al.: 'Towards an interoperability roadmap for the energy transition', e&i Elektrotechnik und Informationstechnik, 2023, 140, pp. 478–487

[4] Dänekas C., González, J.M.: 'Requirements Engineering for Smart Grids', *Standardization in Smart Grids*, Berlin, Heidelberg: Springer Berlin Heidelberg, 2013, pp. 15–37

[5] 'Interoperability Network for the Energy Transition (int:net)', https://intnet.eu/, accessed 16 July 2024

[6] Jimeno, J., Santos Mugica, M., Cortes Borray, A.F., et al.: 'Deliverable D1.1 Repository of interoperability initiatives', (The int:net Consortium, 2022)

[7] Jimeno, J., Santos Mugicia, M., Maqueda, E., et al.: 'Deliverable D1.2 Report on identified interoperability use cases, requirements in the value chain and business models', (The int:net Consortium, 2024)

[8] Dognini, A., Monti, A., Kung, A., et al., 'Blueprint of the Common European Energy Data Space,'' (The int:net Consortium, 2024), https://doi.org/10.5281/zenodo.10964387

[9] 'Find out more about IEC 61850', https://iec61850.dvl.iec.ch, accessed 16 July 2024

[10] Papaioannou, I., Andreadou, N., and Tarramera Gisbert, A.: 'Energy Smart Appliances' Interoperability: Analysis on Data Exchange from State-of-the-art Use Cases', (European Commission - Joint Research Center, 2022), https://doi.org/10.2760/199155

[11] IEC 61970:2024 SER: 'Energy management system application program interface (EMS-API) - ALL PARTS', 2024

[12] Widl, E., Strasser, T., Mac Gregor, C.A., et al.: 'Deliverable D3.1 Testing concepts and procedures harmonisation report', (The int:net Consortium, 2024)

[13] 'CIM Conformity and Interoperability', https://www.entsoe.eu/data/cim/cim-conformity-and-interoperability, accessed 5 July 2024

[14] 'CIM/CGMES IOP', (The int:net Consortium, 2023)

[15] 'Smart Grid Design of Interoperability Tests (SG-DoIT)', https://ses.jrc.ec.europa.eu/sgdoit, accessed 18 July 2024

[16] 'Code of Conduct for Energy Smart Appliances', https://ses.jrc.ec.europa.eu/development-of-policy-proposals-for-energy-smart-appliances, accessed 18 July 2024

[17] 'Interoperability Test Bed', https://joinup.ec.europa.eu/collection/interoperability-test-bed-repository/solution/interoperability-test-bed, accessed 18 July 2024

[18] 'The IES - Approach', https://www.smartgrids.at/interoperabilitaet/ies-initiative, accessed 18 July 2024

[19] 'Living Lab Cologne', https://www.livinglabcologne.com, accessed 18 July 2024

[20] 'Holistic Test Description (HTD)', https://github.com/ERIGrid2/holistic-test-description, accessed 18 July 2024

[21] Raussi, P., Kamsamrong, J., Paspatis, A., et al.: 'Energy Systems Test Case Discovery Enabled by Test Case Profile and Repository', Open Source Modelling and Simulation of Energy Systems (OSMSES), Aachen, Germany, 2023, pp. 1–6